\documentclass{svproc}
\usepackage{graphicx}%
\usepackage{multirow}%
\usepackage{amsmath,amssymb,amsfonts}%
\usepackage{mathrsfs}%
\usepackage[title]{appendix}%
\usepackage{xcolor}%
\usepackage{textcomp}%
\usepackage{manyfoot}%
\usepackage{booktabs}%
\usepackage{algorithm}%
\usepackage{algorithmicx}%
\usepackage{algpseudocode}%
\usepackage{listings}%

\usepackage{array}

\usepackage{url}

\def\orcidID#1{\unskip$^{[#1]}$} % added MR 2018-03-10

\begin{document}
\mainmatter              % start of a contribution
\title{Echoes of Automation: How Bots Shaped Political Discourse in Brazil, 2018-2022}
\titlerunning{Bots and Political Discourse in Brazil, 2018-2022}  % abbreviated title (for running head)
\author{Merve Ipek Bal \and Diogo Pacheco\orcidID{0000-0002-8199-585X}
}
%
% \authorrunning{Author1Name Author1Surname et al.} % abbreviated author list (for running head)
%

\institute{Computer Science Department, University of Exeter, UK\\
\email{\{mib208, d.pacheco\}@exeter.ac.uk}}

\maketitle              % typeset the title of the contribution

\begin{abstract}
In an era where social media platforms are central to political communication, the activity of bots raises pressing concerns about amplification, manipulation, and misinformation. Drawing on more than 315 million tweets posted from August 2018 to June 2022, we examine behavioural patterns, sentiment dynamics, and thematic focus of bot- versus human-generated content spanning the 2018 Brazilian presidential election and the lead-up to the 2022 contest. Our analysis shows that bots relied disproportionately on retweets and replies, with reply activity spiking after the 2018 election, suggesting tactics of conversational infiltration and amplification. Sentiment analysis indicates that bots maintained a narrower emotional tone, in contrast to humans, whose sentiment fluctuated more strongly with political events. Topic modelling further reveals bots’ repetitive, Bolsonaro-centric messaging, while human users engaged with a broader range of candidates, civic concerns, and personal reflections. These findings underscore bots’ role as amplifiers of narrow agendas and their potential to distort online political discourse.
% We would like to encourage you to list your keywords within
% the abstract section using the \keywords{...} command.
\end{abstract}

\section{Introduction}\label{sec1}

Over the past decade, social media platforms have become central to political communication, providing immediacy and wide reach for the dissemination of messaging and public debate~\cite{pepitone2010twitter}. Platforms such as Twitter have been increasingly leveraged by politicians for election campaigns~\cite{ricard2020misinformation}, while automated accounts, commonly known as social bots~\cite{ferrara2016rise}, have emerged as influential actors capable of amplifying narratives, spreading misinformation, and subtly shaping voter sentiment~\cite{pacheco2024bots,shao2017spread,hagen2022}. Network structure further mediates the impact of these actors, and more segregated networks can significantly amplify the spread of misinformation, even when only a minority of actors actively promote it~\cite{karimi2024modelling}. Moreover, the political leanings strongly influence the information exposure of individuals, with conservative accounts clustering in denser networks and encountering more low-credibility content~\cite{chen2021neutral}. These findings highlight the interplay between automated activity, network topology, and partisan exposure, providing important context for studying bot-driven amplification in political discourse.

The 2018 Brazilian presidential election marked a historic turning point, as the first time the country elected a far-right candidate since redemocratization. Twitter and WhatsApp played a central role in shaping candidate visibility and public discourse against a backdrop of political scandal and widespread public mistrust~\cite{recuero2021discursive,evangelista2019whatsapp}. Far-right candidate Jair Bolsonaro effectively leveraged these platforms to engage a highly polarised audience, relying on online messaging amplified by coordinated networks and low-reputation outlets, creating conditions conducive to misinformation, emotional manipulation, and strategic narrative amplification~\cite{teixeira2019polls,Aruguete02012021}.

This study examines bot influence on Brazilian political discourse over nearly the entire presidential term following the 2018 election, from August 2018 to June 2022. The dataset captures a wide range of political events and controversies during this period, including the COVID-19 pandemic, major policy debates, and recurring election-related discussions, providing a comprehensive view of both bot and human activity on Twitter. Extending prior work by Pacheco et al.~\cite{pacheco2024bots}, we conduct a deeper analysis of bot activity across this long-term timeline, focusing on differences in engagement strategies, emotional tone, and thematic content between bot- and human-generated discourse.

To structure our analysis, we address three primary research questions:

\begin{enumerate}
    \item Which Twitter bot behaviours are most prevalent throughout the presidential term, and how do they differ from human activity?
    \item How does the sentiment of political discourse differ between bot-generated and human-generated tweets over time and across key events?
    \item What are the differences in topic themes between bot-generated and human-generated tweets during election controversies and other politically significant periods?
\end{enumerate}

Our analysis shows clear behavioural contrasts. Humans consistently rely on retweets, whereas bots, particularly those identified at higher confidence thresholds, engage more through replies, suggesting deliberate tactics to infiltrate and potentially disrupt conversations.  

This difference extends to emotional tone: bots display a narrower and more stable sentiment range than humans, particularly at lower classification thresholds, indicative of limited emotional variability and a potentially scripted communication style.  

Topic modelling further reinforces this pattern. Bots predominantly amplify a narrow set of Bolsonaro-centric and emotive themes, whereas humans engage with a broader spectrum of candidates, civic issues, and personal reflections, reflecting a more diverse and decentralised discourse.

\section{Methods}
This study analyses bot behaviour and influence in Brazilian political discourse on Twitter over a four-year period (August 30, 2018–June 30, 2022). The dataset, originally collected via the Twitter Streaming API as part of prior work~\cite{pacheco2024bots}, contains more than 315 million tweets. The collection relied on keyword tracking of candidate names, official accounts, campaign hashtags, and electoral authorities, ensuring comprehensive coverage of election-related discussions across both the 2018 and 2022 (partial) electoral cycles.

\subsection{Bot Detection and Dataset Integration}
Raw Twitter JSON files were merged with an existing dataset containing precomputed BotometerLite botscores~\cite{pacheco2024dataset,Yang_Varol_Hui_Menczer_2020}. These scores range from 0 to 1, with higher values indicating greater similarity to automated behaviour. Merging was performed on tweet IDs, yielding a dataset annotated with bot-likelihood scores. For analysis, three thresholds (0.5, 0.7, 0.9) were applied to categorise accounts as bots (score $\geq$ threshold) or humans (score $<$ threshold).

\subsection{Text Preprocessing}
To ensure linguistic quality, tweets were filtered to include only Portuguese content and subjected to a multi-step preprocessing pipeline. Noise elements such as URLs, retweet tags, user mentions, hashtags, emojis, and special characters were removed. Text was normalised to lowercase and stopwords eliminated using the Portuguese stopword list from NLTK~\cite{bird2009nltk}. Quoted text was excluded to focus exclusively on users’ own contributions.

\subsection{Research Question 1: Bot Behaviour During Elections}
To address \textit{“Which Twitter bot behaviours are most prevalent during elections, and how do they differ from humans?”}, a temporal analysis of tweet types was conducted. Tweets were categorised as:
\begin{itemize}
  \item \textbf{Original tweets}: posts that are not replies, retweets, or quotes
  \item \textbf{Retweets}: direct reposts of another user’s tweet
  \item \textbf{Replies}: responses to another user
  \item \textbf{Quotes}: retweets with added commentary
\end{itemize}

For each day, the proportion of tweet types was calculated separately for bots and humans. Percentages normalised activity levels across groups and enabled direct comparison of behavioural trends. To reduce short-term volatility, 14-day centred moving averages were computed. These trends were visualised in side-by-side plots, displaying both raw daily values (thin lines) and smoothed trajectories (bold lines). The 2018 presidential election day (28 October) was marked on all plots for reference.

\subsection{Research Question 2: Sentiment Dynamics}
To address \textit{“How does the sentiment of political discourse differ between bot-generated and human-generated tweets?”}, sentiment scores were computed using a lexicon-based approach. Each tweet was matched against SentiLex-PT02~\cite{sentilex_pt02_2017}, a Portuguese sentiment dictionary, and the raw polarity score was normalised with:
\[
S = \frac{x}{\sqrt{x^2 + \alpha}}, \quad \alpha = 15
\]
where \(x\) is the raw sentiment score and \(S\) is bounded in \([-1, 1]\). Daily mean sentiment values were then calculated for bots and humans and smoothed with a 14-day centred moving average. This enabled the identification of broader sentiment trends across electoral periods.

\subsection{Research Question 3: Thematic Focus}
To answer \textit{“What are the differences in topic themes between bot-generated and human-generated tweets during election controversies?”}, topic modelling was applied to the most politically intense period of the 2018 election (October 1–31). Tweets were split into bot and human corpora using a 0.5 threshold and processed with CountVectorizer from sklearn, applying Portuguese stopwords and filtering terms with \texttt{min\_df=10} and \texttt{max\_df=0.9}.  

Separate Latent Dirichlet Allocation (LDA)~\cite{sievert-shirley-2014-ldavis} models were trained for bots and humans. After testing topic numbers from 5 to 30, twenty topics were selected as the most interpretable. Each topic was represented by its top 15 words, extracted and saved for analysis.

To aid interpretation, the lists of salient terms were collaboratively contextualised using ChatGPT, which assisted in mapping keywords to Brazilian political discourse. Topics were then manually labelled based on recurring terms and thematic coherence. Interactive visualisations were produced using pyLDAvis, allowing inspection of topic separability and word relevance. Salient terms were also ranked by probability weight to enable side-by-side comparison of bot and human topic models, highlighting thematic differences in discourse framing.

\section{Results}
\subsection{Which Twitter bot behaviours are most prevalent during elections, and how do they differ from human activity?} \label{q1}
\begin{figure}[t]

    \centering
    \includegraphics[width=1.0\textwidth]{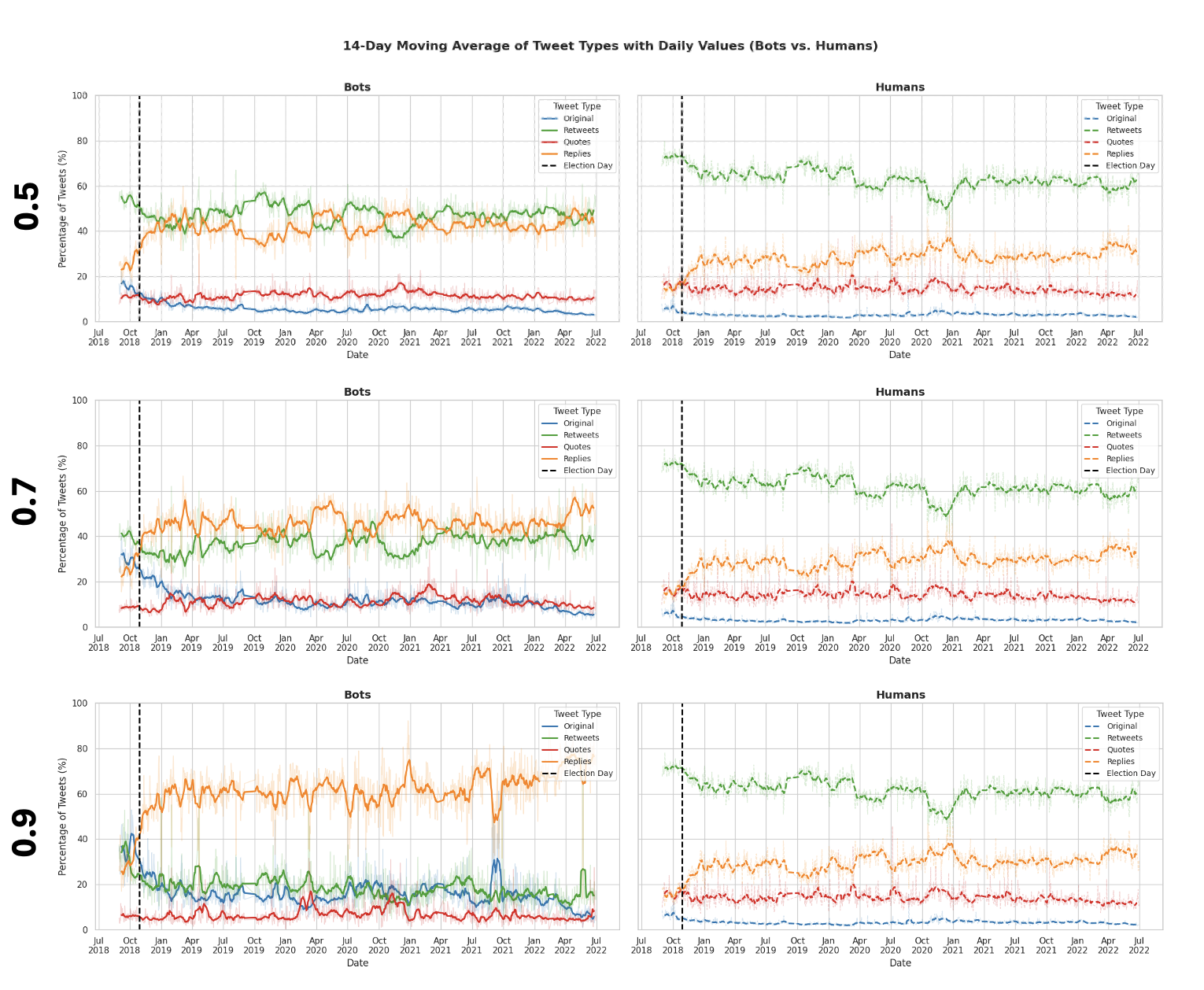}           
    \caption{Fourteen-day moving average sentiment scores for tweets classified as bot- or human-generated, using three BotometerLite thresholds (0.5, 0.7, 0.9) to define bot likelihood. Each panel compares sentiment over time for bots (red) and humans (blue), with shaded lines indicating daily values and solid lines showing smoothed trends. The dashed vertical line denotes the 2018 Brazilian presidential election.}

    \label{fig:tweet_types}

\end{figure}

During the election period, bot activity was primarily dominated by retweets and replies. This distinction becomes especially pronounced at the 0.9 threshold, where the proportion of replies from bot accounts spikes dramatically from approximately 25\% to nearly 80\% following the election. In contrast, the reply rate for human users remains relatively stable, fluctuating between 20\% and 40\% across all thresholds. This sharp increase in bot replies, particularly concentrated in the weeks surrounding the election, suggests the tactical deployment of bots to engage in conversational threads, potentially to amplify messages, disrupt dialogue, or insert propaganda. Such behaviour aligns with patterns seen in coordinated influence operations, where bots exploit reply functions to increase visibility and infiltration.

Original tweets by bots, conversely, decline markedly around the election period and remain consistently low throughout the timeframe. For thresholds 0.5 and 0.7, original tweets account for only 5–10\% of bot activity, and even at the stricter 0.9 threshold, they rarely exceed 20\%, apart from a brief anomaly in September 2021. This limited generation of original content underscores the role of bots as amplifiers rather than originators, focused more on redistributing existing material than contributing novel discourse.

Quote tweets among bots exhibit a minor but stable presence across all thresholds, typically remaining below 10\%, indicating limited contextual engagement. Retweets continue to comprise a large share of bot activity, reinforcing their function in message diffusion rather than dialogue or debate.

Overall, the temporal dynamics of tweet types reveal that bot accounts are highly responsive to political events, with notable shifts in behaviour around the election period. While humans maintain a consistent pattern of retweeting, bots increasingly shift toward replying and therefore disrupting conversations at higher thresholds of classification.
This reactive behaviour, coupled with a persistent lack of original content, indicates that bots function less as independent voices and more as tools for targeted amplification and disruption.

\subsection{How does the sentiment of political discourse differ between bot-generated and human-generated tweets?} \label{q2}

At most thresholds, human sentiment displays greater emotional variability, fluctuating in response to major political events such as elections and controversies. Bot sentiment, in contrast, appears notably more stable and neutral, especially in the smoothed trend lines. This difference suggests that humans engage with political discourse more reactively, while bots maintain a narrower emotional range.

However, it is important to note that these trends represent averaged sentiment scores across large volumes of bot-authored tweets. As such, even if certain bot tweets express highly emotional or polarising content, their overall impact on the group’s average sentiment may be diluted due to the volume and repetitiveness of neutral or mildly worded tweets. Bots often rely on retweets or templated messages, which may lack emotional variability, thereby pulling the average sentiment closer to neutral—even in the presence of sharp, targeted messaging.

\begin{figure}[t]
    \centering
    \includegraphics[width=1.0\textwidth]{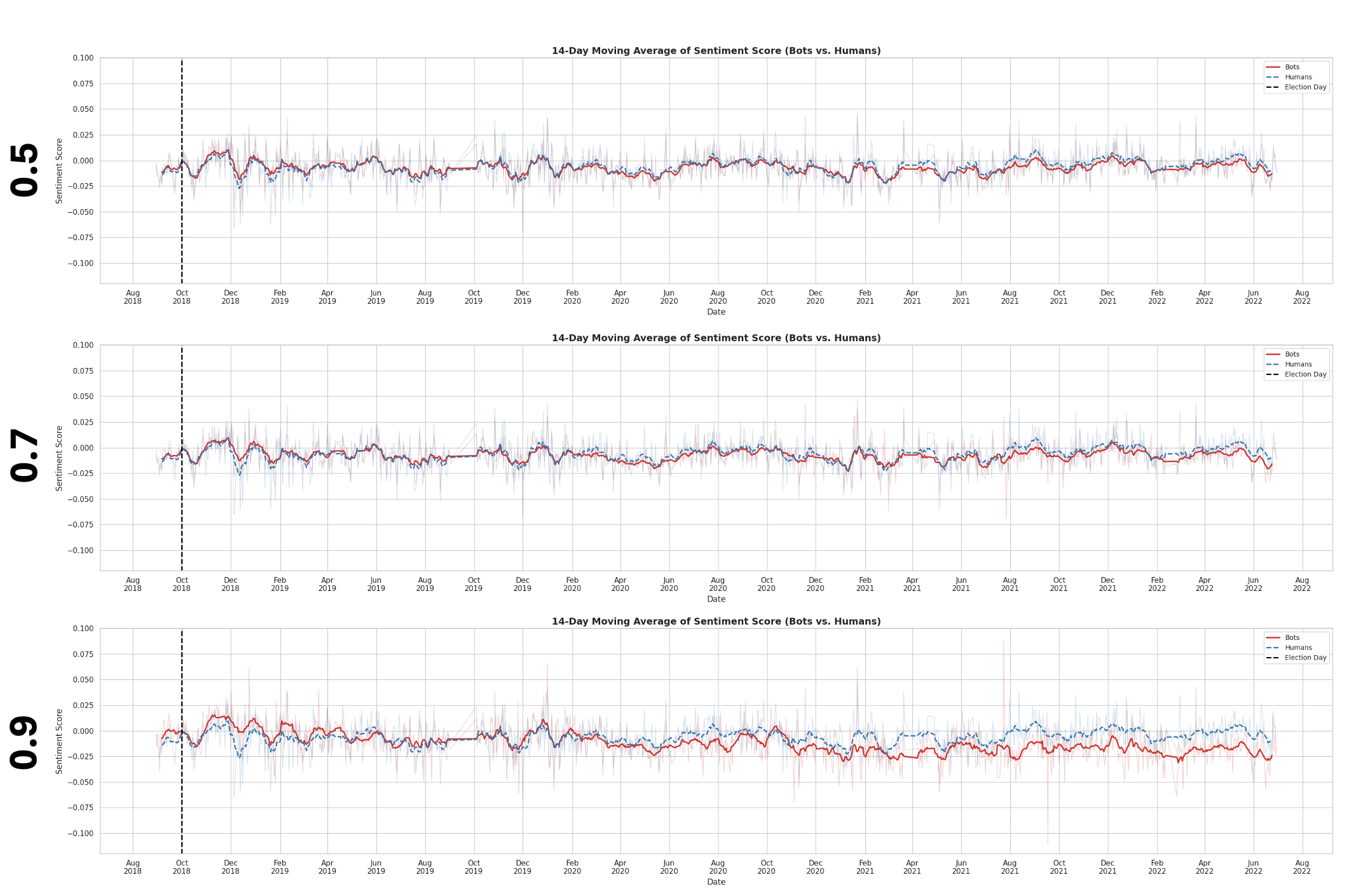}           
    \caption{14-day moving average sentiment scores for tweets classified as either bot- or human-generated, using three different BotometerLite thresholds (0.5, 0.7, 0.9) to define bot likelihood. Each panel compares the average sentiment over time for bots (red) and humans (blue), with shaded lines representing daily fluctuations and solid lines showing smoothed trends. The dashed vertical line marks the 2018 Brazilian presidential election.}
    \label{fig:tweet_sentiments}
\end{figure}

At the 0.5 threshold, where the bot classification is more inclusive, bots exhibit mildly negative but relatively flat sentiment over time. Their sentiment line remains closer to neutral than that of humans, particularly after October 2019, indicating less reactivity to political flashpoints.

At 0.7, which represents a more refined subset of likely bots, the divergence becomes more pronounced. Bot sentiment remains consistently stable, while human sentiment continues to swing more sharply, especially around political milestones. This suggests bots are not just less reactive but possibly constrained by their algorithmic structure, resulting in limited emotional variability.

However, at the 0.9 threshold, the behaviour shifts. Bots begin to show more extreme sentiment spikes than humans. While the mean sentiment still hovers near neutral, the amplitude of the spikes increases, indicating that these high-confidence bots may be more specialised or aggressive in emotional messaging. This higher volatility contrasts with earlier thresholds and may reflect a more strategic or provocative communication style, possibly designed to polarise or provoke engagement.

\subsection{What are the differences in topic themes between bot-generated and human-generated tweets during election controversies?}

To examine differences in political discourse during the 2018 Brazilian election, Latent Dirichlet Allocation (LDA) models with 20 topics were trained separately on bot-generated and human-generated tweets. For brevity, only the first five interpreted topics for each group are presented below, with broader thematic comparisons discussed in the text. Complete topic listings are omitted due to space constraints.

\begin{table}[t]
\centering
\caption{First 5 Interpreted Themes from Bot-Generated Topics (LDA Model)~\cite{chatgpt25}}
\label{tab:bot_lda_topic_interpretation}
\begin{tabular}{p{0.3cm} >{\raggedright\arraybackslash}p{2.5cm} >{\raggedright\arraybackslash}p{8cm}}
\hline
\textbf{\#} & \textbf{Topic Title} & \textbf{Top Words (with translation/explanation)} \\
\hline
1 & Criticism of PT and Institutions & 
pouco (little), pt (Workers’ Party), sim (yes), falta (lack/absence), vergonha (shame), presidente (president), eleição (election), população (population), emprego (job/employment) \\

2 & Haddad and Campaign Messaging & 
haddad (PT candidate), fernando (Fernando Haddad), brasil (Brazil), história (history), sociais (social), redes (networks, e.g., social media), campanha (campaign), democracia (democracy) \\

3 & Anti-PT and Moral Panic & 
pt (Workers’ Party), nunca (never), olha (look/see), gay (gay, used in moral panic debates), kit (“kit gay”, a controversial alleged education material), contra (against), lava (short for Lava Jato, major corruption probe), jato (operation “Car Wash”), bolsonaro (right-wing candidate), mentiras (lies) \\

4 & Religious and Emotional \quad\quad Expressions & 
deus (God), povo (people/nation), forte (strong), abraço (hug, often as a sign-off), amor (love), brasil (Brazil), trabalho (work), kkkk (Brazilian equivalent of ``lol'', laughter) \\

5 & Bolsonaro's Social Media Presence & 
bolsonaro, jair (Jair Bolsonaro), facebook, live (live video broadcast), tse (Supreme Electoral Court), campanha (campaign), candidato (candidate), amanhã (tomorrow) \\
\hline
\end{tabular}
\end{table}

\begin{table}[t]
\centering
\caption{First 5 Interpreted Themes from Human-Generated Topics (LDA Model)~\cite{chatgpt25}}
\label{tab:human_lda_topic_interpretation}
\begin{tabular}{p{0.3cm} >{\raggedright\arraybackslash}p{2.3cm} >{\raggedright\arraybackslash}p{8cm}}
\hline
\textbf{\#} & \textbf{Topic Title} & \textbf{Top Words (with translation/explanation)} \\

\hline
1 & Campaign Dynamics and Public Figures & 
brasil (Brazil), pt (Workers’ Party), nome (name/nominee), bolsonaro, moro (Sérgio Moro, judge linked to anti-corruption cases), haddad (Fernando Haddad, PT candidate), campanha (campaign), justiça (justice) \\

2 & Lula vs. Bolsonaro Narrative & 
haddad, bolsonaro, lula (Luiz Inácio Lula da Silva, former PT president), democracia (democracy), corrupção (corruption), país (country), poste (“post”, pejorative term used to call Haddad Lula’s puppet/placeholder) \\

3 & Voting Process and Turnout & 
turno (round of voting — Brazil has two-round elections), voto (vote), eleição (election), votar (to vote), presidente (president), vamos (let’s/we will), brasil (Brazil) \\

4 & Unity and Celebration & 
brasil (Brazil), obrigado (thank you), juntos (together), parabéns (congratulations), apoio (support), acima (above, e.g. “Brazil above all”), mudar (change) \\

5 & Disillusionment and Candidates & 
marina (Marina Silva, candidate), silva (Marina Silva’s surname), deus (God), governo (government), plano (plan), candidato (candidate), perdeu (lost) \\
\hline
\end{tabular}
\end{table}

The LDA results highlight a stark contrast in topical focus between bots and humans. Bot-generated discourse was dominated by references to Jair Bolsonaro across multiple topics (e.g., Topics 5 and 12 in the full model), often repeating his name alongside terms such as \textit{presidente}, \textit{eleito}, and \textit{campanha}. This repetition suggests a strategy of amplification and agenda-setting, reinforcing a narrow candidate-centric narrative. Bots also engaged in moral panic (Topic 3) and religious appeals (Topic 4), alongside limited multilingual outreach in English and Spanish, pointing to attempts at broader influence. Overall, bot discourse exhibited thematic redundancy and limited lexical diversity.

By contrast, human-generated discourse reflected broader engagement with the electoral process. While Bolsonaro was an important reference point (Topic 2), human users also discussed Haddad, Lula, Ciro Gomes, Marina Silva, and others. Themes spanned voting logistics (Topic 3), ethical concerns (Topic 9 in the full model), socioeconomic issues (Topic 13), and personal reflections on democratic participation (Topic 17). Emotional expression was also more varied, with optimism (Topic 4: “Unity and Celebration”) coexisting with disillusionment (Topic 5). This indicates a more heterogeneous and decentralised debate, where humans deliberated across multiple ideological and civic dimensions rather than amplifying a single narrative.

When aggregating both groups by electoral round, topics shifted over time. In the first round, discourse featured comparisons between candidates (e.g., Ciro, Alckmin, Haddad) and concerns about fraud and misinformation. By the second round, themes became more polarised, with sharper divides between pro- and anti-Bolsonaro messaging and stronger ideological framing. Bots consistently emphasised Bolsonaro’s candidacy, while humans continued to deliberate on voting decisions, socioeconomic issues, and democratic values.

In sum, the analysis shows that while bots and humans occupied similar LDA “spaces,” their substantive contributions diverged. Bots functioned primarily as amplifiers, reinforcing Bolsonaro’s image and agenda through repetition, emotional appeals, and multilingual messaging. Humans, on the other hand, demonstrated broader political engagement, addressing multiple candidates, civic concerns, and personal reflections on democracy. These differences underscore the role of bots in narrowing discourse versus humans in expanding it.

\section{Discussions and Conclusion}
This research offers a comprehensive examination of how automated accounts (bots) engaged in political discourse during the 2018 Brazilian elections on Twitter. Analysing sentiment, engagement patterns, and topic modelling sheds light on the strategic use of emotional tone, content repetition, and network interaction by bots to amplify messages and potentially disrupt online conversations.

One of the central findings is that bots acted primarily as amplifiers rather than originators of content. Compared to human users, bots relied heavily on retweets and replies, particularly increasing reply activity after the election period, suggesting a strategy of embedding messages within existing conversations to boost visibility. This pattern aligns with coordinated inauthentic behaviour observed in other electoral contexts.

Sentiment analysis further revealed that while both bots and humans engaged in largely neutral to negative discourse, human-generated tweets displayed a wider emotional range. Human sentiment exhibited sharper fluctuations around political flashpoints, whereas bot sentiment remained consistently narrower and more neutral. This suggests that bots may struggle to authentically mimic the emotional complexity and variability found in human conversation, likely due to their algorithmic scripting or limited affective capabilities.

Topic modelling uncovered important differences in the thematic focus of bots versus humans. Bots exhibited a repetitive, narrowly defined vocabulary, with much of their discourse revolving around Jair Bolsonaro and related campaign messaging. In contrast, human discourse was broader and more diverse, covering a wider array of candidates, political issues, and emotional expressions, including ethical concerns, voter reflection, and socio-economic debates. This indicates that while bots may contribute to message amplification, humans drive a more complex, multifaceted political dialogue.

Interestingly, the distinction between bot and human behaviour was at times subtle. This may be attributed to the limitations of the BotometerLite classifier, which, while effective, may misclassify sophisticated bots that imitate human-like patterns. It also highlights a broader concern: as artificial intelligence advances, bots are increasingly able to mimic authentic human discourse, blurring the line between organic and synthetic political communication.

These findings underscore the growing importance of digital literacy and critical engagement on social media. As misinformation and artificial amplification contribute to political polarisation, it is crucial for users to remain vigilant, fact-check information, and cultivate a reflective approach to online discourse. In an era where bots can shape conversations and reinforce ideological divides, fostering an informed digital public becomes not just a personal responsibility but a democratic imperative.

\subsection*{Declaration}

For the purpose of open access, the authors have applied a Creative Commons Attribution (CC BY) licence to any accepted manuscript version arising from this submission.

\bibliographystyle{splncs04}
\bibliography{references} % Entries are in the references.bib file

@article{chen2021neutral,
  title={Neutral bots probe political bias on social media},
  author={Chen, Wen and Pacheco, Diogo and Yang, Kai-Cheng and Menczer, Filippo},
  journal={Nature communications},
  volume={12},
  number={1},
  pages={5580},
  year={2021},
  publisher={Nature Publishing Group UK London}
}

@inproceedings{karimi2024modelling,
  title={Modelling Misinformation Spread: The Role of Network Density in Diverse Social Structures},
  author={Karimi, S and Oliveira, M and Pacheco, D},
  year={2024},
  booktitle={Social Simulation Conference (SSC 2024)},
  publisher={European Social Simulation Association (ESSA)}
}

@article{evangelista2019whatsapp,
  title={WhatsApp and political instability in Brazil: targeted messages and political radicalisation},
  author={Evangelista, Rafael and Bruno, Fernanda},
  journal={Internet policy review},
  volume={8},
  number={4},
  pages={1--23},
  year={2019},
  publisher={Berlin: Alexander von Humboldt Institute for Internet and Society}
}

@article{recuero2021discursive,
  title={Discursive strategies for disinformation on WhatsApp and Twitter during the 2018 Brazilian presidential election},
  author={Recuero, Raquel and Soares, Felipe Bonow and Vinhas, Ot{\'a}vio and others},
  journal={First Monday},
  volume={26},
  number={1},
  year={2021},
  publisher={University of Illinois at Chicago}
}

@article{ferrara2016rise,
  title={The rise of social bots},
  author={Ferrara, Emilio and Varol, Onur and Davis, Clayton and Menczer, Filippo and Flammini, Alessandro},
  journal={Communications of the ACM},
  volume={59},
  number={7},
  pages={96--104},
  year={2016},
  publisher={ACM New York, NY, USA}
}

@inproceedings{pacheco2024bots,
author = {Pacheco, Diogo},
title = {Bots, Elections, and Controversies: Twitter Insights from Brazil's Polarised Elections},
year = {2024},
isbn = {9798400701719},
publisher = {Association for Computing Machinery},
address = {New York, NY, USA},
url = {https://doi.org/10.1145/3589334.3645651},
doi = {10.1145/3589334.3645651},
booktitle = {Proceedings of the ACM on Web Conference 2024},
pages = {2651–2659},
numpages = {9},
location = {, Singapore, Singapore, },
series = {WWW '24}
}

@misc{pacheco2024dataset,
  author = {Pacheco, Diogo},
  title = {Dataset for the paper - Bots, Elections, and Controversies: Twitter Insights from Brazil’s Polarised Elections},
  year = {2024},
  howpublished = {\url{https://doi.org/10.5281/zenodo.10669936}}
}

@article{Yang_Varol_Hui_Menczer_2020, title={Scalable and Generalizable Social Bot Detection through Data Selection}, volume={34}, url={https://ojs.aaai.org/index.php/AAAI/article/view/5460}, DOI={10.1609/aaai.v34i01.5460}, abstractNote={&lt;p&gt;Efficient and reliable social bot classification is crucial for detecting information manipulation on social media. Despite rapid development, state-of-the-art bot detection models still face generalization and scalability challenges, which greatly limit their applications. In this paper we propose a framework that uses minimal account metadata, enabling efficient analysis that scales up to handle the full stream of public tweets of Twitter in real time. To ensure model accuracy, we build a rich collection of labeled datasets for training and validation. We deploy a strict validation system so that model performance on unseen datasets is also optimized, in addition to traditional cross-validation. We find that strategically selecting a subset of training data yields better model accuracy and generalization than exhaustively training on all available data. Thanks to the simplicity of the proposed model, its logic can be interpreted to provide insights into social bot characteristics.&lt;/p&gt;}, number={01}, journal={Proceedings of the AAAI Conference on Artificial Intelligence}, author={Yang, Kai-Cheng and Varol, Onur and Hui, Pik-Mai and Menczer, Filippo}, year={2020}, month={Apr.}, pages={1096-1103} }

@inproceedings{bird2009nltk,
  title     = {Natural Language Toolkit (NLTK)},
  author    = {Bird, Steven and Klein, Ewan and Loper, Edward},
  booktitle = {Proceedings of the ACL 2009 Conference: Short Papers},
  pages     = {19--22},
  year      = {2009},
  organization = {Association for Computational Linguistics}
}

@misc{sentilex_pt02_2017,
  author       = {Silva, Jorge and Oliveira, Sara},
  title        = {SentiLex-PT02},
  year         = {2017},
  publisher    = {CLUL - Centro de Linguística da Universidade de Lisboa},
  url          = {https://b2find.eudat.eu/dataset/b6bd16c2-a8ab-598f-be41-1e7aeecd60d3},
  note         = {Accessed: 2025-04-12}
}

@inproceedings{sievert-shirley-2014-ldavis,
    title = "{LDA}vis: A method for visualizing and interpreting topics",
    author = "Sievert, Carson  and
      Shirley, Kenneth",
    editor = "Chuang, Jason  and
      Green, Spence  and
      Hearst, Marti  and
      Heer, Jeffrey  and
      Koehn, Philipp",
    booktitle = "Proceedings of the Workshop on Interactive Language Learning, Visualization, and Interfaces",
    month = jun,
    year = "2014",
    address = "Baltimore, Maryland, USA",
    publisher = "Association for Computational Linguistics",
    url = "https://aclanthology.org/W14-3110/",
    doi = "10.3115/v1/W14-3110",
    pages = "63--70"
}

@misc{pepitone2010twitter,
  author = {Pepitone, Julianne},
  title = {Twitter users not so social after all},
  year = {2010},
  month = {March},
  journal = {CNN Money},
  note = {Available: \url{https://money.cnn.com/2010/03/10/technology/twitter\_users\_active/index.htm?hpt=Mid}}
}

@article{ricard2020misinformation,
  author = {Ricard, Julie and Medeiros, Juliano},
  title = {Using misinformation as a political weapon: COVID-19 and Bolsonaro in Brazil},
  journal = {Harvard Kennedy School Misinformation Review},
  volume = {1},
  number = {3},
  year = {2020},
  note = {Available: \url{https://misinforeview.hks.harvard.edu}}
}

@article{shao2017spread,
  title={The spread of fake news by social bots},
  author={Shao, Chengcheng and Ciampaglia, Giovanni Luca and Varol, Onur and Flammini, Alessandro and Menczer, Filippo},
  journal={arXiv preprint arXiv:1707.07592},
  volume={96},
  number={104},
  pages={14},
  year={2017}
}

@misc{chatgpt25,
    author = {ChatGPT},
    title = {Come up with topic titles, for the following topics and terms in the context of 2018 Brazilian elections: 
    Topic 1 terms:pouco, pt, sim, falta, vergonha, presidente, nao, vamos, pois, país, eleição, população, eua, emprego, histórica...},
    year = {2025},
}

@article{hagen2022,
author = {Loni Hagen and Stephen Neely and Thomas E. Keller and Ryan Scharf and Fatima Espinoza Vasquez},
title ={Rise of the Machines? Examining the Influence of Social Bots on a Political Discussion Network},
journal = {Social Science Computer Review},
volume = {40},
number = {2},
pages = {264-287},
year = {2022},
doi = {10.1177/0894439320908190},
URL = { 
    
        https://doi.org/10.1177/0894439320908190
},
eprint = { 
    
        https://doi.org/10.1177/0894439320908190
}
}

@inproceedings{teixeira2019polls,
  title={Polls, plans and tweets: an analysis of the candidates’ discourses during the 2018 brazilian presidential election},
  author={Teixeira, Carlos and Kurtz, Gabriela and Leuck, Lorenzo and Sanvido, Pedro and Scherer, Joana and Tietzmann, Roberto and Manssour, Isabel and Silveira, Milene},
  booktitle={Proceedings of the 20th Annual international conference on digital government research},
  pages={439--444},
  year={2019}
}

@article{Aruguete02012021,
author = {Natalia Aruguete and Ernesto Calvo and Tiago Ventura and},
title = {News Sharing, Gatekeeping, and Polarization: A Study of the \#Bolsonaro Election},
journal = {Digital Journalism},
volume = {9},
number = {1},
pages = {1--23},
year = {2021},
publisher = {Routledge},
doi = {10.1080/21670811.2020.1852094},
URL = { 
    
        https://doi.org/10.1080/21670811.2020.1852094

},
eprint = { 
        https://doi.org/10.1080/21670811.2020.1852094

}

}
\end{document}